# Maintenance-free Operation of WDM Quantum Key Distribution System through a Field Fiber over 30 Days

Ken-ichiro Yoshino,[1,*] Takao Ochi,[1] Mikio Fujiwara,[2]
Masahide Sasaki,[2] and Akio Tajima[1]

[1]*NEC Corporation, 1753 Shimonumabe, Nakahara-ku, Kawasaki, Japan*
[2]*National Institute of Information and Communications Technology, 4-2-1 Nukuikitamachi, Koganei, Tokyo, Japan*
[*]*yoshino@bp.jp.nec.com*

**Abstract:** Maintenance-free wavelength-division-multiplexing quantum key distribution for 30 days was achieved through a 22-km field fiber. Using polarization-independent interferometers and stabilization techniques, we attained a quantum bit error rate as low as 1.70% and a key rate as high as 229.8 kbps, making the record of total secure key of 595.6 Gbits accumulated over an uninterrupted operation period.

## 1. Introduction

Contemporary cryptographic protocols are implemented by computer algorithms based on that certain mathematical problems are practically impossible to solve using current computer resources and well-known attacks. This kind of algorithmic cryptography is not provably secure, and vulnerable to off-line attacks that can occur long after a secret message has been sent. A well known scheme of provably secure protocol is Vernam's one-time pad (OTP), in which a truly random key whose length is the same as that of a plain text should be shared between the legitimate sender and receiver for encryption and decryption simply by exclusive OR with the plain text, and is used only once. Quantum key distribution (QKD) can provide a means to share such a truly random key on demand through a physical channel of photon transmission. Its security can be proven against the eavesdropper with unbounded abilities.

Although QKD technology has become in practical use, its speed and distance are still limited. Commercial products typically operate at key rates of about a few kbps over a few tens of km field fibers [1, 2]. Therefore they are often used not in the OTP mode but in hybrid mode combined with contemporary cryptographic protocols, such as refreshing the seed keys for symmetric key encryption scheme. As for QKD systems in laboratories, key rates reach a few hundred kbps over 40~50 km field fibers [3–5], which enable one to support real-time OTP encryption of video data. Recently wavelength-division multiplexing (WDM) QKD system is developed [6-7], providing a flexible solution to support variety of applications including secure voice transmission and real-time secure TV conferencing with OTP encryption. In fact, increasing the key rate only with single-channel QKD has a bottleneck due to a speed limit of single-photon detectors, whose novel types operate in a GHz range [8-11], but further improvement is not easy.

Now a current important task for putting such state-of-the-art QKD technologies to practical uses is quality assurance, especially for long-term maintenance-free operation. The high-speed QKD specifications mentioned above have been recorded only for two weeks at longest [12]. Their stability has not yet been at a sufficient level to be certified for practical services in commercial environments. The stability also directly affects the certification of QKD security. Experimental guarantee of stable key generation with low enough quantum bit error rate (QBER) for a long period is a prerequisite for quality assurance of a QKD system.

We report on maintenance-free, continuous key generation over 30 days using two quantum channels through a 22-km field fiber. Our WDM-QKD system has polarization-independent interferometers and newly implemented stabilization control, which can optimize various parameters automatically. We achieved stable key generation of higher than 200 kbps in total and QBER of 1.61% and 1.86% for each channel on average for 30 days, and accumulated secure keys of 595.6 Gbits in total over uninterrupted duration.

## 2. WDM-QKD System

In the previous works, we established a compact optical transmission block consisting of planar lightwave circuits (PLCs), which are superior in mass productivity, reliability, polarization-insensitivity, and controllability [13-15]. Figure 1 shows simplified functional diagram of the optical system. In a transmitter (Alice), a PLC-based 2×2 asymmetric Mach-Zehnder interferometer (AMZI) converts an optical pulse from a laser diode (LD) driven at 1.24 GHz into a pair of pulses with a time delay of 400 ps. An encoder puts data and basis information on the pair for preparing the time-bin signal in phase-time coding. Then variable optical attenuator (VOA) attenuates the power of the pair to 0.5 photons. The quantum signal is combined with the clock signal for precise and automatic synchronization by using a WDM coupler and is transmitted through the same fiber. In a receiver (Bob), a PLC-based 2×4 AMZI, which is totally passive, followed by four photon detectors, decodes the time-bin signal. The delay tuning of the AMZIs is made with the temperature controller (TC) within an accuracy of 0.01 K to attain a total extinction ratio as high as 20 dB [13].

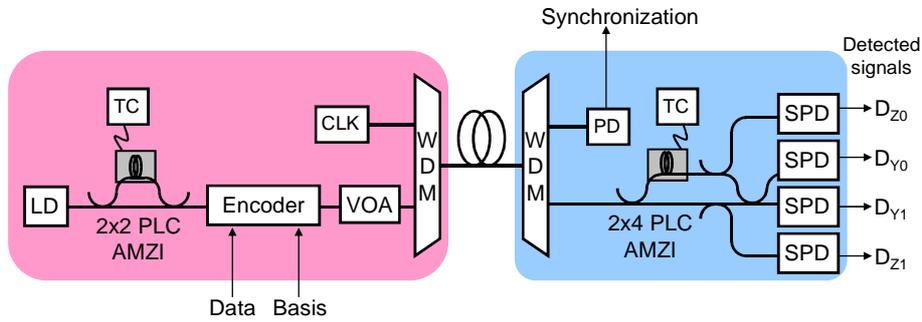

Fig. 1. Block diagram of single-channel QKD. LD: laser diode, PLC: planar lightwave circuit, AMZI: asymmetric Mach-Zehnder interferometer, TC: temperature controller, VOA: variable optical attenuator, CLK: optical clock source, WDM: wavelength-division multiplexing coupler, PD: photo diode, SPD: single photon detector.

These technologies can be optimally used to implement our GHz-clocked WDM-QKD system, which can potentially handle up to eight wavelength channels allocated on the ITU-T 100-GHz grid from 1545.3 to 1550.9 nm. One more important technology is required to make the WDM-QKD system cost-effective, which is a *colorless* interferometer consisting of the PLC AMZI whose temperature is controlled for polarization-independent interference and a technique of phase mismatch compensation for every wavelength. Figure 2 shows our WDM-QKD system. In the transmitter, optical pulses emitted by LDs are multiplexed with a WDM coupler. The multiplexed pulses pass through the PLC AMZI, and are converted into the time-bin signals. They are then demultiplexed, and each wavelength component is encoded with bit data and basis information by phase and amplitude modulation as denoted by $\phi_A(\lambda_n)$. Additional phase modulators denoted by $\Delta\theta(\lambda_n)$ follow to compensate for the phase mismatches. The receiver decodes the signals almost in the reverse manner. Thanks to this colorless interferometer with the PLCs, we could have demonstrated WDM-QKD with reduced system size, cost and control complexity [6-7].

Optical components including colorless interferometer and other elements required by a QKD system such as synchronization and key distillation are mounted onto the advanced telecom computing architecture (ATCA) chassis and 19-inch rack. Our WDM-QKD system is designed for a secure key rate >1 Mbps over a 10-dB loss channel, when eight wavelength channels are fully operated.

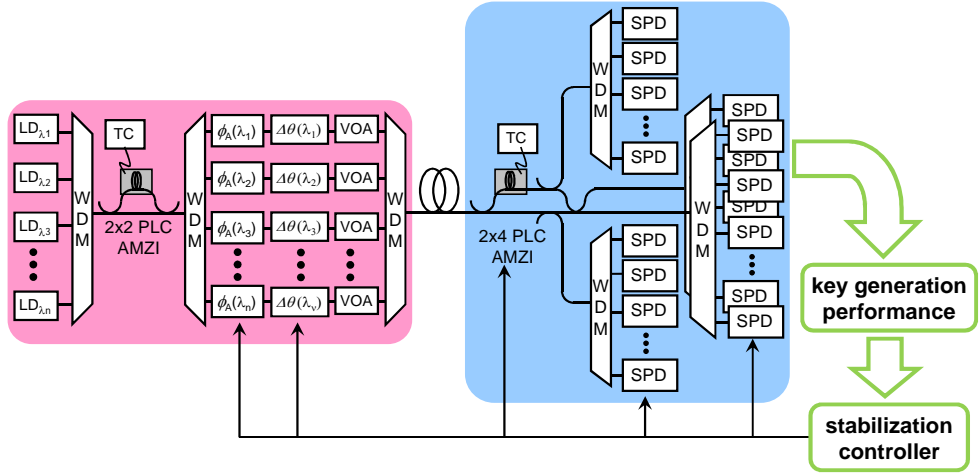

Fig. 2. WDM-QKD system with the colorless interferometer. Synchronization is omitted.

## 3. Stabilization Techniques

Even if we set parameters to the optimal values at the beginning of operation, the key generation performance would degrade after a certain amount of time. This is caused by temperature change of the transmission fiber or location, the drift of power-supply voltage, and temporal change of devices. For long-term reliable operation, we have to suppress these effects. For example, polarization variation in transmission fiber is critical to the extinction ratio of the usual interferometer, and dramatically increases the QBER within a short time. The polarization variation could be solved by the polarization-independent interferometer based on the PLC technique [13] as mentioned above. For another example, fiber expansion delays the optimal detection timing of photon detectors. In our system, the delay of 50 ps causes degradation of photon count rate to around 80%. Therefore we optimize detection timing with an accuracy of 12.5 ps. To improve the reliability of the QKD system, such optimizations have to be done for various components constantly. We have newly added the following stabilization controls :

- Detection timing of the photon detectors
- Bias voltage of the encoding modulator
- Temperature of the PLC interferometer
- Amplitude of the phase compensation modulator

The stabilization control software periodically retrieves the key generation performance (photon count rates and QBERs) from the receiver Bob (Fig. 2). Then it supervises the device components listed above to vary the operating point slightly different from the present ones and monitors the key generation performance again to see the results. Finally the software compares those results on tunings and chooses the optimal parameters to the best operating point. This control scheme keeps each parameter optimal and results in stable key generation performance even if the environmental condition and the component condition fluctuate.

## 4. Long-term Field Demonstration

We demonstrated our 2-channel WDM-QKD for BB84 protocol using wavelengths, $\lambda 1$ (1547.72 nm) and $\lambda 2$ (1550.92 nm). The transmission channel is a 22-km field installed single-mode fiber in a loop-back configuration. More than 95% of the channel is in an aerial fiber over poles. The total loss is 12.6 dB. In the receiver, two avalanche photodiode (APD) systems [11] were used. The quantum efficiency and dark count rate were 10–15% and 1-2 k count/s.

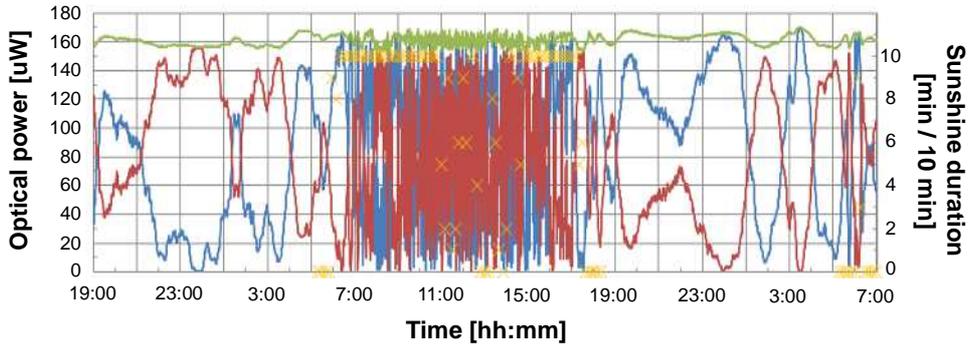

Fig. 3. Polarization stability measurement of 36 hours. The red and blue lines are orthogonal polarizations measured through a polarizing beam splitter. The green line shows the total transmitted power. The yellow crosses denote the sunshine duration for 10 minutes.

Figure 3 shows a typical variation of fiber transmission characteristics in terms of polarization drift. Polarization fluctuates all the time. The fluctuation is more apparent during the day.

Figures 4 (a) and (b) plot the QBERs, the sifted and secure key rates for the two channels for 30 days. The QKD performance of each channel is summarized in Table 1. During 30 days of continuous operation, stable secure key generation at an average rate of 229.8 kbps in total and the averaged QBER of 1.70% were achieved.

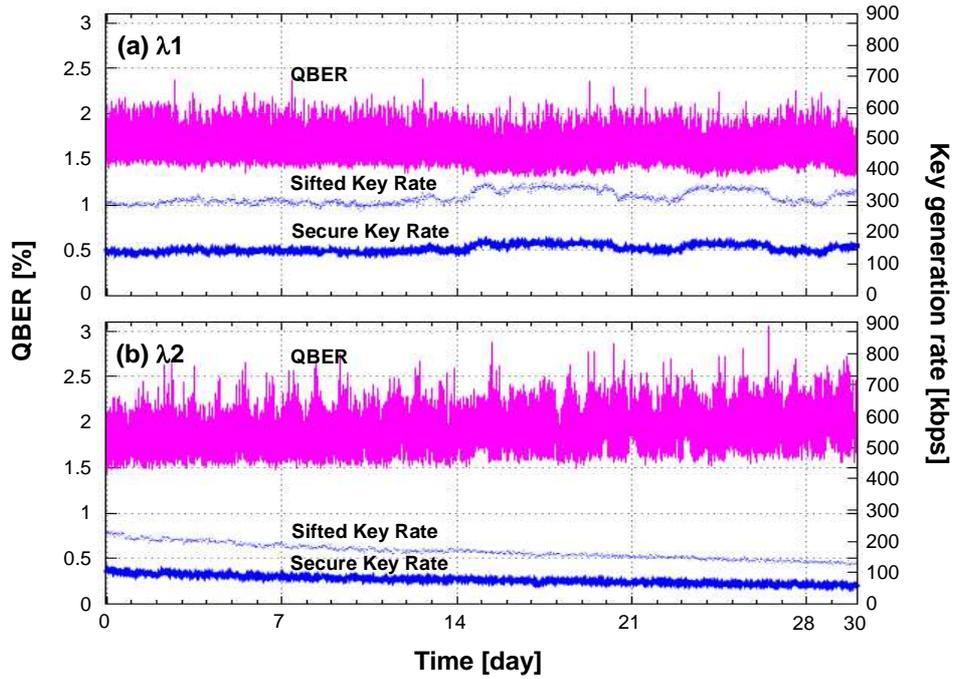

Fig. 4. Key generation performance of $\lambda1$ (a) and $\lambda2$ (b) over 30days.

**Table 1. Key Generation Performance for 30 days**

| Wavelength [nm] | QBER [%] | Sifted key rate [kbps] | Secure key rate [kbps] |
|---|---|---|---|
| λ1: 1547.72 | 1.61 | 315.3 | 151.5 |
| λ2: 1550.92 | 1.86 | 168.0 | 78.3 |
| Total | 1.70 | 483.3 | 229.8 |

In Table 2, we compare some results of recent long-term field operations. The total number of secure bits distributed over the 22 km installed fiber for a 30-day uninterrupted duration of continuous operation amounts to 595.6 Gbits. This is an order of magnitude larger than those in the previous works, such as 70.3 Gbits over the 3.7 km installed fiber for 327 days (SQ3 link from 3 Sep 2009 to 26 July 2010) by Stucki et al. [16], and 63.3 Gbits by Dynes et al. over the 45-km installed fiber for 60 hours [17].

**Table 2. Comparison of Recent Long-term Field Operations**

|  | This work | Swiss Quantum [16] | Toshiba-NICT [17] | SEQURE (CV-QKD) [18] |
|---|---|---|---|---|
| Uninterrupted duration of continuous operation | 30 days | 327 days | 60 hours | 85 days |
| Distance [km] | 22 | 3.7 | 45 | 17.7 |
| Channel loss [dB] | 12.6 | 2.5 | 14.5 | 5.6 |
| Secure key rate [kbps] | 229.8 | 2.5 | 293 | 0.6 |
| Average QBER [%] | 1.70 | 1.2 (estimated from Fig. 7 [16]) | Not available | Not used |
| Total secure keys distributed [Gbits] | 595.6 | 70 | 63.3 | 4.4 |
| Channel loss normalized secure keys [Tbits/channel loss] | 10.8 | 0.12 | 1.78 | 0.016 |

To capture the quality of long-term QKD operation, one may introduce a new figure of merit, namely the channel-loss normalized secure key, defined by an amount of distributed secure key in total divided by a channel loss. This metric shows the size of accumulated secure key over an uninterrupted period through an assumed lossless channel. One can estimate how large secure key can reliably be distributed over a given channel, by multiplying this metric by a real channel loss to be used. Our result is 10.8 Tbits, which is an order of magnitude larger than those ever reported to our knowledge.

**5. Conclusion**
We have developed a high-speed QKD system based on the scalable WDM technique with polarization-independent interferometers and stabilization control of various parameters. By driving the system with two channels, we achieved 30 days of maintenance-free continuous key generation with QBERs of 1.70% and a secure key rate of 229.8 kbps through a 22-km field fiber with 12.6-dB loss. We introduced a new metric, the channel-loss normalized secure key to quantify the quality of long-term QKD operation. Our result amounted to the world record of 10.8 Tbits.

This research is supported by Research and Development of Secure Photonic Network Technologies, the Commissioned Research of National Institute of Information and Communications Technology (NICT), Japan.